\documentclass[aps,prl,twocolumn,showpacs]{revtex4}

\usepackage{amsmath} \usepackage{epsfig} \usepackage{graphics}

 \renewcommand{\Im}{{\rm Im}}
\renewcommand{\Re}{{\rm Re}}

\def\ba{\begin{array}} \def\ea{\end{array}}

\begin{document}


\title{Strong coupling of a qubit to shot noise}

\author{Udo Hartmann} \email[]{udo.hartmann@physik.lmu.de}

\author{Frank K. Wilhelm}
\affiliation{Physics Department, Arnold Sommerfeld Center for Theoretical Physics, and Center for NanoScience, \\
Ludwig-Maximilians-Universit\"at M\"unchen,  Theresienstr. 37, D-80333 M\"unchen, Germany}



\begin{abstract}
We perform a nonperturbative analysis of a charge qubit in a double quantum dot structure
coupled to its detector. We show that strong detector-dot interaction tends to slow down and halt
coherent oscillations. The transitions to a classical and a low-temperature quantum 
overdamping (Zeno) regime are studied. In the latter, the physics of the dissipative phase 
transition competes with the effective shot noise.
\end{abstract}




\maketitle


The study of fluctuations and noise provide deep insights into quantum processes in 
systems with many degrees of freedom. If coupled to a few-level system 
such as a qubit, fluctuations usually lead to destabilization of general
qubit states and induce decoherence and energy relaxation. One important manifestation
is the back-action of detection on qubits \cite{Braginsky}.
This topic has been extensively studied in the regime of weak coupling between
qubit and noise source \cite{Clerk}. It has been shown that the qubit
dephases into a mixture of qubit eigenstates (dephasing), whose classical
probabilities thermalize to the noise temperature at a longer time scale.
Mesoscopic charge detectors such as quantum point
contacts (QPCs) \cite{Elzerman} and radio-frequency single electron transistors (rf-SETs)
\cite{Rimberg}, 
whose low-temperature noise is shot noise \cite{Blanter,Aguado}, are particular powerful
detectors as they provide high resolution \cite{Pilgram} and
potentially reach the quantum limit. 
A particular attractive regime for qubit applications is
the QND regime, realized if the qubit Hamiltonian and the qubit-detector
coupling commute \cite{Braginsky,GurvitzPrager}.

We study a quantum point contact potentially strongly coupled to the coordinate (left/right)
of a double quantum dot charge qubit \cite{Hayashi,Gorman} by a 
nonperturbative approach involving the Gaussian and noninteracting blip approximations. 
We analyze the qubit at the charge degeneracy point, where the two lowest
energy eigenstates are delocalized between the qubits. In the weak coupling
regime, low-temperature relaxation would thus always delocalize charge. 
We show that, in strong coupling, the
qubit state gets localized in one of the dots. 
Localization is manifest by a suppression of both 
the coherent oscillations and the incoherent tunneling rate. 
This ``freezing'' of the state also 
applies a high bias and can {\it e.g.} lock an excited state. Thus, in the 
strong coupling regime, the environment naturally pushes the physics to the 
QND limit even if the bare Hamiltonian is not QND. We point out the analogy of this physics to the case of the dissipative
phase transition in oscillator bath models \cite{Leggett}, which in the QPC competes with the
nonequilibrium induced by the voltage driving the shot noise.

\begin{figure}[h]
\centering
\includegraphics*[width=\columnwidth]{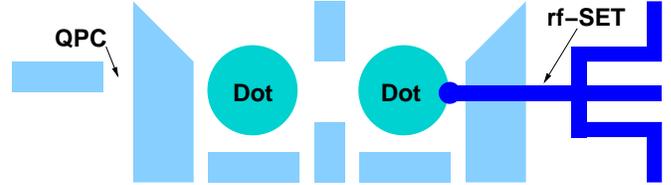}
\caption{Schematic view of the double dot system analyzed
see {\it e.g.} Refs.~[\onlinecite{Elzerman,Rimberg}]. The QPC and rf-SET detectors 
can be used {\em alternatively}, both options are discussed in the paper.} 
\label{fig:dots}
\end{figure}

We consider the case of a degenerate two-state system (TSS), realized by 
the charge states in a double quantum dot structure (see Figure~\ref{fig:dots}). 
These charge can be read out by the current through a nearby quantum point contact.
The Hamiltonian for the TSS with time-dependent fluctuation 
$\tilde{\varepsilon}(t)$ reads
\begin{equation}
H_{\rm sys}= \frac{\hbar}{2}\left( \begin{array}{cc}
\tilde{\varepsilon}(t) & \Delta \\
\Delta & -\tilde{\varepsilon}(t) \\
\end{array}\right)
\rightarrow \tilde{H}_{\rm sys} = \frac{\hbar\Delta}{2}\left( \begin{array}{cc}
0 & e^{i\phi}\\
e^{-i\phi} & 0\\
\end{array}\right).
\label{eq:hamiltonian} 
\end{equation}
In the last step of eq.~(\ref{eq:hamiltonian}), we applied a Polaron 
transformation \cite{Schoen} introducing the fluctuating phase
$\phi = \int^t dt' \tilde{\varepsilon}(t')$, with $\tilde{\varepsilon}(t)=\varepsilon + \delta\varepsilon(t)$,
for the tunneling matrix elements in the qubit. The microscopic foundation of
the noise term $\delta\varepsilon (t)$ for a QPC is given in Refs.~[\onlinecite{Levinson,Aguado}]
and for an SET in Refs.~[\onlinecite{Korotkov,Aassime,Johansson,Kaeck}].

Without loss of generality, we assume $\langle\hat{\sigma}_z(0)\rangle=1$. We can 
now formally solve the Liouville equation. The expectation value of $\hat{\sigma}_z$, the difference
of occupation probabilities of the dots, satisfies a closed equation
\begin{eqnarray}
\langle\dot{\hat{\sigma}}_{z}(t)\rangle & = & -\Delta^2 \int_{0}^{t}\limits dt' \cos\left[\varepsilon(t-t')\right]\langle e^{i\delta\phi(t)}e^{-i\delta\phi(t')}\rangle\langle\hat{\sigma}_{z}(t')\rangle \nonumber\\
& = & -\Delta^2 \int_{0}^{t}\limits dt' \cos\left[\varepsilon(t-t')\right]e^{J(t-t^\prime)} \langle\hat{\sigma}_{z}(t')\rangle, \label{eq:sz2}
\end{eqnarray}
where the second line of eq.~(\ref{eq:sz2}) has been derived by assuming that 
the noise represented by $J(t-t^\prime)$ is stationary. 
This procedure is analogous to the noninteracting blip 
approximation (NIBA) of the path-integral 
solution of the Spin-Boson model \cite{Leggett,Weiss}. This automatically
includes a Gaussian approximation to the shot noise \cite{Aguado}.
This approach
is nonperturbative in $\phi$ and a good approximation in the two cases
$\varepsilon=0$ and $|\varepsilon|\gg|\Delta|$. 

We start with the charge-degeneracy case $\varepsilon=0$. Here, we can 
solve eq.~(\ref{eq:sz2}) in Laplace space and find 
\begin{equation}
\mathcal{L}\left[\langle\hat{\sigma}_{z}(t)\rangle\right] = \frac{1}{s+\Xi(s)}\label{eq:laplace}\ ,
\end{equation}
with the Laplace-transformed self-energy
$\Xi(s) = \Delta^2\int_{0}^{\infty}\limits dt e^{-st} e^{J(t)}$.
The phase correlation function $J(t)$ as seen by the dots 
reads \cite{Aguado}
\begin{equation}
J(t) = \frac{2\pi}{\hbar R_K}\int_{-\infty}^{\infty}\limits d\omega \frac{|Z(\omega)|^2}{\omega^2} S_{I}(\omega)\left( e^{i\omega t}-1\right)\ , \label{eq:j1}
\end{equation}
where $S_{I}(\omega)$ is the full current noise in the QPC that for sufficient environmental impedance is 
given \cite{Aguado} by
\begin{eqnarray}
S_{I}(\omega)& = &\frac{4}{R_K}\sum_{m}^{N}\limits D_{m}(1-D_{m})\Bigg\{\frac{\hbar\omega+eV}{1-e^{\beta(\hbar\omega+eV)}}+ \nonumber\\
&{}& + \frac{\hbar\omega-eV}{1-e^{-\beta(\hbar\omega-eV)}}\Bigg\} + \frac{4}{R_K} \sum_{m}^{N} D_{m}^2 \frac{2\hbar\omega}{1-e^{-\beta\hbar\omega}}\nonumber \label{eq:si1}\\
\end{eqnarray}
and the transimpedance $Z(\omega)$ between
qubit and point contact. In eq.~(\ref{eq:si1}), $V$ is the bias voltage of the QPC, $R_K$ 
is the quantum resistance, and $D_m$ is the transmission eigenvalue of the $m$th conductance channel. 

{\em Semiclassical limit:}
We now discuss the resulting dynamics in a number of limiting cases.
We start by first taking the limit
$\omega\rightarrow 0$. This corresponds to $\hbar\Delta,\hbar\varepsilon\ll eV,k_BT$, 
{\it i.e.} the qubit probes the shot noise at energy scales much lower than its internal
ones. Here, the noise expression [eq.~(\ref{eq:si1})]
becomes frequency independent \cite{Blanter}. We can then compute the semiclassical 
spectral function $J_c(t)=-\gamma_c t$.
Here, we have assumed a frequency-independent transimpedance controlled 
by a dimensionless parameter $\kappa$, $|Z(\omega)|^2\approx \kappa^2 R_K^2$
and $\gamma_c=2\pi^2\kappa^2 R_K S_I(0)$ with
$S_{I}(0) = \frac{4}{R_K}\sum_{m}^{N}\limits D_{m}(1-D_{m}) eV 
\coth\left(\frac{\beta eV}{2}\right) + \frac{4}{R_K}\sum_{m}^{N}\limits D_{m}^2 \frac{2}{\beta}.$
The self-energy is then readily calculated and analytical, so we can
go back from Laplace to real time and obtain
\begin{equation}
\langle\hat{\sigma}_{z}(t)\rangle = \Bigg[ \cos\left(\omega_{\rm eff,c}t\right)+ \frac{\gamma_c}{2\omega_{\rm eff,c}}\sin\left(\omega_{\rm eff,c}t\right)\Bigg]e^{-\frac{\gamma_c}{2}t}, \label{eq:sz3}
\end{equation}
where $\omega_{\rm eff,c} = \sqrt{\Delta^2-\frac{\gamma_c^2}{4}}$. 
We observe that the coherent oscillations of the qubit decay on a scale 
$\gamma_c^{-1}$ and get slowed down. At $\gamma_c=2\Delta$, the damping becomes
critical and the oscillations disappear, ending up with a purely exponential
overdamped regime at $\gamma_c>2\Delta$. This crossover corresponds to the 
classical overdamping of a harmonic oscillator.  
Even in the overdamped regime, the qubit decays exponentially to $\langle\hat{\sigma}_z(t)\rangle\rightarrow 0$ at 
long times, {\it e.g.} it gets completely mixed by the shot noise, whose noise
temperature is high $k_B T_{\rm noise}\simeq {\rm max}\lbrace eV,k_B T\rbrace \gg 
\hbar\Delta$. Note that it is possible 
to discuss the overdamped regime, where 
$\gamma_c$ is {\em not} a small parameter and our theory is also non-Markovian, see eq.~(\ref{eq:sz2}),
capturing the necessary time-correlations arising in strong coupling. 

\begin{figure}[h]
\centering
\includegraphics*[width=\columnwidth]{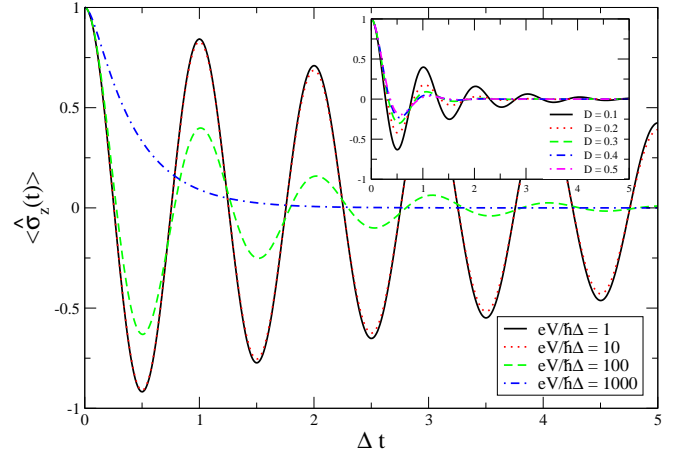}
\caption{Semiclassical limit: expectation value
$\langle\hat{\sigma}_{z}(t)\rangle$ as a function of 
time and varied bias voltages $V$ at $\varepsilon=0$. The other parameters are $T = 0.1$~K, $D = 0.1$,
$\Delta = 1.524\cdot 10^9$~1/s, and $\kappa = 0.02$. Inset: as a function of time and the QPC 
transmission $D$ with fixed QPC bias voltage $eV = 100\ \hbar\Delta$.} 
\label{fig:plot1}
\end{figure}
Figure~\ref{fig:plot1} shows the resulting dynamics in the one-channel case. With increasing bias voltage $V$ over the QPC, 
the expectation value $\langle\hat{\sigma}_{z}(t)\rangle$ drops down quite fast.
The transmission $D$ of the QPC has also
an important impact on the stability of the oscillations of 
$\langle\hat{\sigma}_{z}(t)\rangle$ (see inset of Figure~\ref{fig:plot1}). 
At $D = 0.5$, the expression for $S_I(0)$ has 
a maximum, therefore the oscillations are there maximally suppressed.
$S_I(0)$ represents the shot noise of the QPC in the low frequency regime \cite{Balestro}.
The more noise the QPC provides, the quicker the oscillations decay. Note that changes in the QPC
transmissions (and therefore the Fano factor) do not play any role other than entering the total noise level.

{\em Quantum limit:} Now, we let $T\rightarrow 0$ and leave $\omega$ arbitrary.
$S_{I}(\omega)$ reads in this limit
\begin{eqnarray}
S_{I}(\omega) & = & \frac{4}{R_K}\Bigg[ \sum_{m}^{N}\limits D_{m}(1-D_{m})\Big\{\left(\hbar\omega+eV\right)\theta(\hbar\omega+eV)+ \nonumber \\
&& +\left(\hbar\omega-eV\right)\theta(\hbar\omega-eV)\Big\}+\sum_{m}^{N}\limits D_{m}^2 2\hbar\omega \theta(\hbar\omega)\Bigg] \label{eq:si4}.\nonumber\\
\end{eqnarray}

This shape is dominated by two terms, which resemble the Ohmic spectrum at
low $T$, $S_{\Omega}\propto \omega \theta (\omega)$ with shifted 
origins of energy.
For computing the quantum correlation function $J_q(t)$, an ultraviolet 
cutoff $\omega_c$ has to be introduced, which
physically originates either from the finite bandwidth of the electronic
bands in the microscopic Hamiltonian or from the high-frequency limitations
of the transimpedance $Z(\omega)$. We end up with the long-time limit for
$J_q(t)$ applicable at $\hbar\Delta\ll eV$
\begin{equation}
J_q(t) = -\alpha_1+\alpha_2 \ln\left[ \left(\frac{eV}{\hbar}\right)^F\frac{1}{\omega_c}\ t^{F-1}\right]-\gamma_q t + i\alpha_3 \label{eq:j3}.
\end{equation}
This holds for any number of channels, for simplicity we 
concentrate on the single-channel case with a  Fano factor
then is given by $F=1-D$, which we use from now on. Here, we can introduce 
$\alpha_2=g=16\pi\kappa^2 D$, the dimensionless conductance as seen by
the qubit, $\alpha_1=g\gamma D$, $\alpha_3=\pi g/2$ and $\gamma_q=\pi g(1-D)eV/2\hbar$. The resulting
self-energy is now non-analytical
\begin{equation}
\Xi(s) = \Delta^2_{\rm eff}\frac{\left(s+\gamma_q\right)^{gD-1}}{\left(\frac{eV}{\hbar}\right)^{gD}}\ e^{i\pi g/2},
\end{equation}
where we have introduced the effective tunnel splitting
$\Delta_{\rm eff}^2= \Delta^2 e^{-\gamma gD}\left(\frac{eV}{\hbar\omega_c}\right)^g\Gamma\left(-gD + 1\right)$.
In our regime, $\omega_c\gg eV/\hbar \gg 1/t\simeq \Delta$, this 
expression resembles the renormalized $\Delta$ of the Spin-Boson model 
\cite{Weiss} and we have $\Delta_{\rm eff}\ll\Delta.$ This is a sign of massive entanglement between system and
detector \cite{Costi,Jordan}. Note that similar to the adiabatic scaling treatment in Ref.~[\onlinecite{Leggett}],
the NIBA is compatible with forming entangled
states between system and bath. This has been numerically confirmed,
for the Spin-Boson model, in Ref.~[\onlinecite{Costi}]. An elegant approach to this
system reflecting entanglement and use of the measurement result in the perturbative regime 
has been given in Ref.~[\onlinecite{Milburn}]. The main difference 
in our shot noise case is that 
the infrared cutoff entering the renormalization and controlling 
the final expressions appears to be $V$ instead of $\Delta$. In particular, $\Delta_{\rm eff}$ grows
with $eV$, which indicates that nonequilibrium shot noise competes with the Spin-Boson-like suppression.

The self-energy is analytical only at $F=1$, which corresponds to the no-noise case
$D=0$. 
Due to the generally non-analytic self-energy, it is difficult to compute the full 
real-time dynamics by back-transformation to the time domain. The structure of the result will
be $\left\langle\hat{\sigma}_z(t)\right\rangle=P_{\rm cut}(t)+P_{\rm coh}(t)+P_{\rm incoh}(t)$ 
\cite{Weiss}. 
For our case of $\varepsilon=0$, there is no incoherent exponential decay $P_{\rm incoh}$. $P_{\rm cut}$
is a nonexponential branch cut contribution.
In the following, we concentrate on the coherent part $P_{\rm coh}(t)$, given through the poles
$s_i=-\gamma_{\rm eff}\pm i\omega_{\rm eff}$ of $\Xi$ with finite imaginary part, and hence this
leads to damped harmonic oscillations with frequency $\omega_{\rm eff}$ and decay rate $\gamma_{\rm eff}$. 

\begin{figure}[h]
\centering
\includegraphics*[width=\columnwidth]{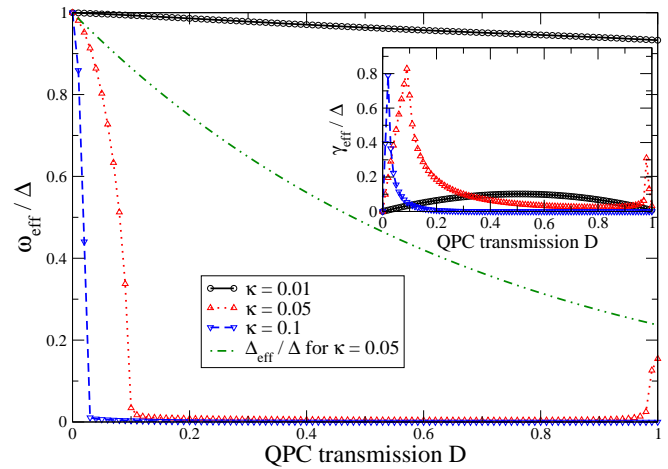}
\caption{Quantum limit: imaginary parts of the numerically determined poles 
as a function of the QPC transmission $D$. The other parameters are 
$eV = 10^{2}\ \hbar\Delta$, $\omega_c = 10^{12}\ \Delta$. 
Inset: real parts of the poles as a function of the QPC transmission $D$.} 
\label{fig:plot2}
\end{figure}

Close to $D=0$, we can characterize these poles perturbatively. We
find a renormalized oscillation frequency $\omega_{\rm eff}$, namely
$\omega_{\rm eff} = \Re \left(\sqrt{\Delta_p^2\left(1+\frac{i\pi}{2}g\right)-\frac{\gamma_q^2}{4}}\right)$ 
whereas $\gamma_{\rm eff}= \frac{\gamma_q}{2} \mp \Im\left(\sqrt{\Delta_p^2\left(1+\frac{i\pi}{2}g\right)-\frac{\gamma_q^2}{4}}\right) $.
Here, $\Delta_p^2$ is defined as $\Delta_p^2 = \Delta^2\left(1+g \ln\left(\frac{eV}{\hbar\omega_c}\right)\right)$.
For arbitrary $F$ or $D$, we can solve the pole equation numerically, see Fig.~\ref{fig:plot2}. 
With the numerical results from Figure~\ref{fig:plot2}, one can again calculate the 
Laplace back-transformation, where the two residues of the kind
$a_{-1} = \frac{e^{s_i t}(s_i+\gamma_q)}{s_i(2-gD)+\gamma_q}$
have to be summed up. This leads finally again to decaying oscillations as already mentioned above.

We see that at sufficiently strong coupling to the detector, a finite Fano factor can lead to 
a complete suppression of the coherent oscillations, whereas the decay rate
increases. Both these tendencies {\em together} show that a finite Fano factor brings the system 
closer to charge localization.
In fact, for sufficient damping, we can tune the tunneling frequency all the way to zero by increasing $D$. 
On the other hand, also $\gamma_{\rm eff}$ can become very small --- in these points the detector
completely localizes the particle up to nonexponential contributions.  At other values of $D$, unlike
the dissipative phase transition in the Spin-Boson model, the hot electrons driving the shot noise
again drive the relaxation rate close to its bare value, and thus this resembles the classical overdamping
case.

\begin{figure}[h]
\centering
\includegraphics*[width=\columnwidth]{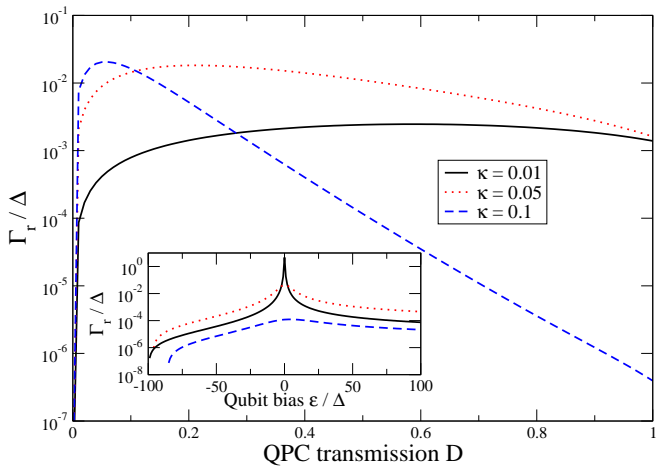}
\caption{Quantum limit: relaxation rate $\Gamma_{r}$ 
as a function of the QPC transmission $D$. The other parameters are 
$\varepsilon = 10\ \Delta$, $\Delta = 1.524\cdot 10^9$~1/s, 
$eV = 10^{2}\ \hbar\Delta$, $\omega_c = 10^{12}\ \Delta$. 
Inset: relaxation rate $\Gamma_{r}$ as a function of the
qubit bias $\varepsilon$. Other parameters as above, but with $D=0.5$.} 
\label{fig:plot3}
\end{figure}

This scenario is not limited to $\varepsilon=0$. NIBA permits to reliably 
study the opposite regime $\varepsilon\gg\Delta$ as well. 
As already shown in Refs.~[\onlinecite{Leggett,ChemPhys}], the resulting dynamics
is dominated by incoherent exponential relaxation dominating over $P_{\rm coh}$ and $P_{\rm cut}$. The relaxation rate is
\begin{equation}
\Gamma_{r} = 2 \Re \left[\Xi(i\varepsilon+0)\right]=2\Delta^2_{\rm eff}
\Re\left[\frac{(i\varepsilon + \gamma_q)^{gD-1}}{(eV)^{gD}}\ e^{i\pi g/2}\right]\label{eq:gammar}.
\end{equation}
This again demonstrates the slowdown (through $\Delta_{\rm eff}$) of the decay to the other dot due
to the interaction with the detector. Notably, this rate does not display standard detailed balance 
at $T=0$, rather, around $\varepsilon=0$, the rate is smeared out on a scale of $\gamma_q$, reflecting
the role of the nonequilibrium shot noise temperature. 
We have plotted this result in Figure ~\ref{fig:plot3}.

Another view on this is that the effective
size of the noncommuting term between qubit and detector, given by $\Delta_{\rm eff}$, is 
reduced, hence the strong interaction brings the effective Hamiltonian closer
to a QND situation.

On the other hand, such dynamics is known
as the quantum Zeno effect. Note that unlike standard derivations\cite{Braginsky,Gurvitz,GurvitzPrager}, this has been derived in a 
nonperturbative way, which is consistent with the necessary strong coupling and which
retains the non-Markovian structure. 

Summarizing the QPC results, we can observe that, on the one hand, the system shows
traces of the physics of environment-induced localization, which competes with classical overdamping
by effectively "hot" electrons at finite voltage and
somewhat reinforced at finite Fano factor. This can be understood as follows: the dissipative phase
transition occurs when the environmental noise is highly asymmetric in frequency and when the full
bandwidth plays a role. At high voltage, the asymmetry of the shot noise spectrum is
reduced \cite{Aguado}. In fact, the $\gamma_q t$ contribution in the correlation function $J_q(t)$ 
resembles the finite temperature term in the correlation function of the Ohmic Spin-Boson model
--- both terms originate from the zero-frequency part of the noise. 

A similar analysis on back-action by strong coupling of a QPC to a quantum device --- there an Aharonov-Bohm experiment \cite{Buks} ---
has been done in Ref.~[\onlinecite{Aleiner}]. That work concentrates on a stationary 
situation and weak hopping into the dot, whereas in our case the dots are not connected to 
leads. The inter-dot interaction however is strong and we concentrate on the real-time dynamics. 

These results can be extended to shot noise sources other than 
QPCs. In fact, it may today be quite challenging to reach $\kappa$-values
high enough, such that slowdown and localization can be observed, 
when the noise source has only a few open channels. An attractive alternative
is given by readout using {\em metallic} SETs fabricated on another
sample layer \cite{Rimberg}, see Fig.~\ref{fig:dots}.
In these devices, there is a number of rather opaque conductance channels.

In that case, we use the expression \cite{Korotkov,Aassime,Johansson,Kaeck} 
of the voltage noise of the SET (only valid for small frequencies)
\begin{equation}
S_V(\omega,\omega_I) = 4 \frac{E_{\rm SET}^2}{e^2}\frac{4\omega_I}{\omega^2 + 16\omega_I^2}, \label{eq:sets}
\end{equation}
where $E_{\rm SET}= \frac{e^2}{2 C_{\rm SET}}$ is the charging energy of the SET
and $\omega_I = I/e$ is the tunneling rate through the SET. 
Then the final result for $\langle \hat{\sigma}_z(t)\rangle$ is again the same as 
in eq.(\ref{eq:sz3}). The difference, of course, is that $\gamma_c$ is now defined
as $\gamma_c = \frac{2 \pi^2 \kappa^2 E_{\rm SET}^2}{\hbar R_K e^2 \omega_I}$.
The full quantum mechanical analysis in the low-temperature regime works
along the same lines as the QPC case but goes beyond the scope of this Letter.

We performed a nonperturbative analysis of the quantum 
dynamics of a double quantum dot coupled to shot noise. We analyze the crossover
from under- to overdamped oscillations in the classical case. In the quantum
case, we show that at strong coupling the oscillations show the same behavior, competing
with a critical slowdown similar to the dissipative phase transition. This can 
be interpreted as the onset of a Zeno effect. 

We thank M.J.~Storcz, M.~Sindel, L.~Borda, A.~K\"ack, J.~von Delft,
L.P.~Kouwenhoven, and U.~Weiss for clarifying discussions. This work was supported by DFG
through SFB~631 and in part by the National
Security Agency (NSA) and Advanced Research and Development Activity
(ARDA) under Army Research Office (ARO) contract number P-43385-PH-QC.

\end{document}